\documentclass[apjl]{emulateapj}

\usepackage{natbib}
\usepackage{graphicx}
\usepackage{rotating}
\usepackage{epsfig}
\usepackage{epstopdf}
\usepackage{enumitem}

\newcommand{\Rs}{$ R_{\odot}$}
\newcommand{\de}{$^{\circ}$}

\newcommand{\kms}{km$\,s^{-1}$}
\newcommand{\thh}{$^{th}$}
\newcommand{\Ha}{H$\alpha$}

\begin{document}

\shorttitle{A solar tornado observed by AIA/SDO} \shortauthors{Li,
Morgan, Leonard and Jeska}
\title{A solar tornado observed by AIA/SDO: Rotational flow and evolution of magnetic helicity
in a prominence and cavity}
\author{ Xing Li\altaffilmark{1}, Huw Morgan\altaffilmark{1,2}, \& Drew Leonard\altaffilmark{1},
Lauren Jeska\altaffilmark{1}} \affil{\altaffilmark{1}Sefydliad
Mathemateg a Ffiseg, Prifysgol Aberystwyth, Ceredigion, Cymru, SY23
3BZ} \affil{\altaffilmark{2}Institute for Astronomy, University of
Hawaii, 2680 Woodlawn Drive, Honolulu, HI 96822, USA}
\email{xxl@aber.ac.uk}

\begin{abstract}
During 2011/09/24, as observed by the Atmospheric Imaging Assembly
(AIA) instrument of the Solar Dynamic Observatory (SDO) and
ground-based \Ha\ telescopes, a prominence and associated cavity
appeared above the southwest limb. On 2011/09/25 8:00UT material
flows upwards from the prominence core along a narrow loop-like
structure, accompanied by a rise ($\geq$50,000km) of the prominence
core and the loop. As the loop fades by 10:00, small blobs and
streaks of varying brightness rotate around the top part of the
prominence and cavity, mimicking a cyclone. The most intense and
coherent rotation lasts for over three hours, with emission in both
hot ($\sim$1MK) and cold (hydrogen and helium) lines. We suggest
that the cyclonic appearance and overall evolution of the structure
can be interpreted in terms of the expansion of helical structures
into the cavity, and the movement of plasma along helical structures
which appears as a rotation when viewed along the helix axis. The
coordinated movement of material between prominence and cavity
suggest that they are structurally linked. Complexity is great due
to the combined effect of these actions and the line-of-sight
integration through the structure which contains tangled fields.
\end{abstract}

\maketitle

\section{Introduction}
\label{intro}

Filaments are highly complicated magnetic structures which lie in
the lowest corona. Their structure and dynamics at small and large
scales is not yet fully explained. Recent extensive reviews of their
composition, structure and dynamics are given by
\citet{labrosse2010} and \citet{mackay2010}. The relation of a
filament to surrounding magnetic structure is also complicated.
Viewed above the limb, a quiescent prominence will often be situated
within, or at the base of, a large system of magnetic loops.
Observed in white light and in the extreme ultraviolet (EUV), a
semicircular or circular region of closed loops surrounding the
prominence is relatively dark compared to the surrounding corona,
and is therefore labeled a coronal cavity \citep{waldmeier1970,
gibson2010, reeves2012}. It has been shown, however, that cavities
are not depleted of density, but are at a very high temperature on
the order of 2MK \citep{habbal2010shrouds}. The relation between the
cavity and filament is still unclear. The general filament/cavity
model is of an arcade of loops anchored at the photosphere with the
filament constrained within the loop system. The arcade can be
raised above the photosphere in the form of a helical flux rope
\citep{low1995}.

Reports of long-lived rotation, or cyclonic, behaviour of
non-eruptive prominences are sporadic but have been made  for a long
time. Such phenomena are called `tornadoes' due to their appearance
but their physics are of course very different to that of
terrestrial tornadoes. \citet{pettit1925} describes in detail the
behaviour of prominences and categorize some as `tornado/spiral'.
\citet{ohman1969} measured lineshifts of H$\alpha$ for filaments on
the disk, and found a shift consistent with rotation of the filament
at velocity $\sim$10\kms. \citet{liggett1984} made a study of 51
prominences and found 5 which showed rotation, with apparent
velocities of 15-75\kms, and interpreted the rotation in terms of a
twisting of magnetic structure, and invoke reconnection as a way to
explain the long-lived rotation. \citet{wang2010} describe
continuous rotational movement of filament cavities observed by the
ExtremeUltraViolet Imaging Telescope (EIT) aboard the Solar and
Heliospheric Observatory (SOHO). This movement is interpreted as a
`pinch-off' of a system of arcade loops surrounding a filament,
leading to a helical flux rope. Flow of material along the original
arcade is then restricted to rotate around the helix.

In this Letter, we report a unique activation  of a quiescent
prominence observed by SDO/AIA \citep{lemen2012}. Such prominences
are known to produce emission at temperatures to about $\log
T(\mbox{K}) \approx 5.5$ and show motions of up to 70km s$^{-1}$
\citep{wang1999,chae2000apj,kucera2003,kucera2006}, and
upward-moving jets may be a mechanism injecting mass into
prominences \citep{chae2000solphys}. We describe the phenomenon in
detail in section \ref{observations}, and give further discussion
and provide possible interpretations in section \ref{discussion}.

\section{Observations}
\label{observations}
The AIA instrument aboard SDO measures EUV light in several narrow
wavelength channels each of which  is dominated by an emission line
formed at a particular temperature \citep{lemen2012}. Its high time
($\sim$12s) and spatial (0.6") resolution provide a new view of the
dynamics of chromospheric and coronal structures. The observations
presented here are mostly of the 171\AA\ bandpass channel, dominated by
emission of Fe$^{8+}$ formed at $\sim$0.7MK, and the 304\AA\ channel
dominated by He$^{1+}$ emission at $10^4$K \citep[e.g.]{odwyer2010}.
The filament under study is at position angle 215\de\
(counter-clockwise from North). Seen in H$\alpha$ in a daily
sequence of Big Bear Solar Observatory (BBSO) observations prior to
the active phase under study, it is a nondescript filament, rather
dim and ill-defined, forming part of a chain of similar filaments at
the same latitude.

\begin{figure*}[h]
\begin{center}
\includegraphics[width=2.8cm,angle=-90]{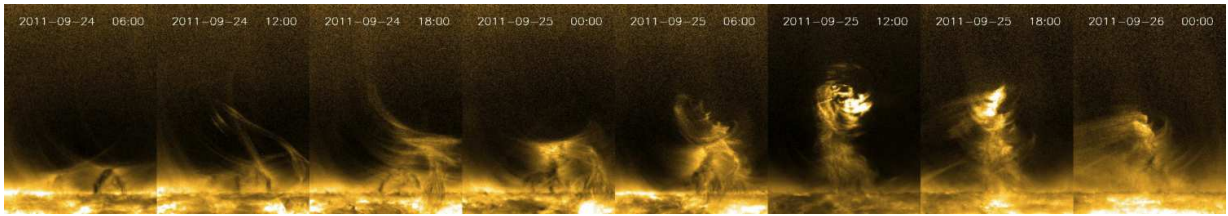}
\end{center}
\caption[]{Development of the prominence tornado structure in the
AIA 171\AA\ channel over  almost two days, in six hour time
increments from 2011/09/24 06:00 (left) to 2012/09/26 00:00 (right).
These images are converted from the original images into polar
coordinates, and show a section of the corona from position angle
210 to 220\de, and height 0.99 to 1.25\Rs.} \label{2dayevolution}
\end{figure*}

\begin{figure*}[h]
\begin{center}
\includegraphics[width=2.8cm,angle=-90]{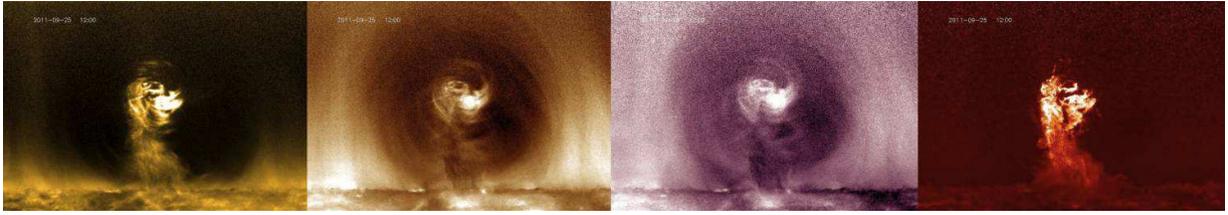}
\end{center}
\caption[]{The different appearance of the tornado in different AIA
wavelength channels  at time 12:00. The main ion contributing to the
signal, approximate wavelength, and approximate peak temperature
sensitivity are from left to right: Fe$^{8+}$ 171\AA\ (0.8MK),
Fe$^{11+}$ 193\AA\ (1.5MK), Fe$^{13+}$ 211\AA\ (1.8MK), and
He$^{1+}$304\AA\ ($10^4$K). The signal is too low in the other
channels to warrant display.} \label{channels}
\end{figure*}

Almost two days of data are analysed from 2011/09/24 06:00 to
2011/09/26 00:00, during which time the  prominence becomes active,
the tornado is formed, and disappears. Fig. \ref{2dayevolution}
shows the development of the whole structure during this period.
What is immediately apparent from this time sequence is the
similarity of structure at 2011/09/24 18:00 and 2011/09/25 00:00
with other studies of filaments and their associated cavities, for
example \citet{regnier2011}. The cavity is suspended above the limb,
and the filament is based directly below the U-shape cradle forming
the base of the cavity. The cavity is most clearly seen in the
193\AA\ and 211\AA\ channels as shown in fig. \ref{channels}, which
are dominated by emission lines at 1.5 and 1.8MK respectively. the
cavity is difficult to see in the 171\AA\ channel, and is invisible
in the 304\AA\ channel. Apparent in fig. \ref{2dayevolution} are
dark barbs, probably components of one of the filament legs, rooted
at the base of the structure from the beginning of the observation
period. This configuration is consistent with the type of model
described by, for example, \citet{vanballegooijen1989} of a large
system of loops or helix enclosing the tighter helix and cool gas of
the filament itself.

From 2011/09/25 00:00 onwards, the evolution of the structure is
considerably different to  that of \citet{regnier2011} where an
eruption of the cavity and filament was observed. Between 2011/09/25
02:30-03:10 the whole structure experiences a large-scale and
short-lived wobble initially towards the pole, and small blobs are
seen to appear and disappear in the cavity immediately above the
filament. Accompanying this movement is a maelstrom of small-scale
activity amongst dark fibrils at the filament base, extending
upwards towards the base of the cavity. By 25/09/25 06:00 (5\thh\
panel of Fig. \ref{2dayevolution}), the filament and cavity have
developed a distinct tornado-like appearance, with a large circular
structure atop a narrower pillar.

At $\sim$8:00, a significant movement of material from the main
body of the filament into the cavity along a very fine channel (the
width is about 3-5 pixels) is observed. By $\sim$8:20, more flow
channels appear and the flows seem to come from both sides of the
prominence. These channels rise and fall back along curved
trajectories, indicating the motions are along curved magnetic field
lines. That the fine channels of flows can break into segments is possibly
due to surface instabilities (Ryutova et al. 2010). The swirling
motions of the channels around the prominence suggest the presence
of helical magnetic fields, but it is difficult to see any helix clearly
before 10:00. These motions make the prominence appear 50,000km
higher than before $\sim$8:00. From $\sim$10:00, there is a large
new injection of material into the filament base from a narrow
channel at one side of the filament. The upflow of this material
towards the cavity base is obvious in the highest time resolution
images (see online animation). The origin of the flow can be traced
to a location at least 14-18Mm above the solar limb.

Following the upflow at 10:00, and for the next $\sim$3 hours, there
is a spectacular series  of movements at the head of the tornado,
with streaks and blobs of varying brightness following circular
paths counterclockwise around the top of the filament pillar - what
was previously a dark cavity. Fig. \ref{rotation} shows some still
images of this action. To truly appreciate the beauty of this event,
the online animation should be viewed. Blobs of material flow into
space which was previously dark, highlighting magnetic structures
which are otherwise invisible. At first (10:00) material is seen
moving along a thin channel and by 10:10 the thin channel is already
widened and a helix-like structure with at least seven turns is very
obvious. The sudden appearance of a similar tightly-wound helix is
repeated again at about 11:00, and a less tightly-wound helix is
apparent at $\sim$11:45. The very core of the tornado head is bright
and complex, with strange slow rotation and movements of filamentary
structure. A bright helix can be identified in the mid-left part of
the structure at 11:45 while the right part shows more tightly wound
structure. At about 12:00, a tangled helix or a group of helices at
the core of the tornado head evolves in a very complicated manner.
The line of sight (LOS) integration, and the complexity of the
structure, prevents any certainty in interpreting this evolution. If
there are two or more helical structures we would expect some
interaction - reconnection possibly, and the development of kink
instabilities which may lead to entanglement of helices (Sakurai,
1976). An event, possibly reconnection, is labelled `V' in fig.
\ref{rotation}. The apparent downflows at 12:30 (labelled `E')
appear to originate from the region where the bright V-shaped loop
system and the loop next to it are seen to come into contact at
$\sim$ 12:00 (see Fig. \ref{rotation}). Of course, the contact could
also be a projection effect.

By 18:00 the head of the tornado has
dimmed, the rotational movement has stopped and by 2011/09/26 00:00
the tornado has disappeared, leaving wispy strands extending at
obtuse angles relative to the radial into the region previously
occupied by the filament pillar. The main period of coherent
rotation lasts for approximately 3 hours.

\begin{figure*}[t]
\includegraphics[width=9.0cm]{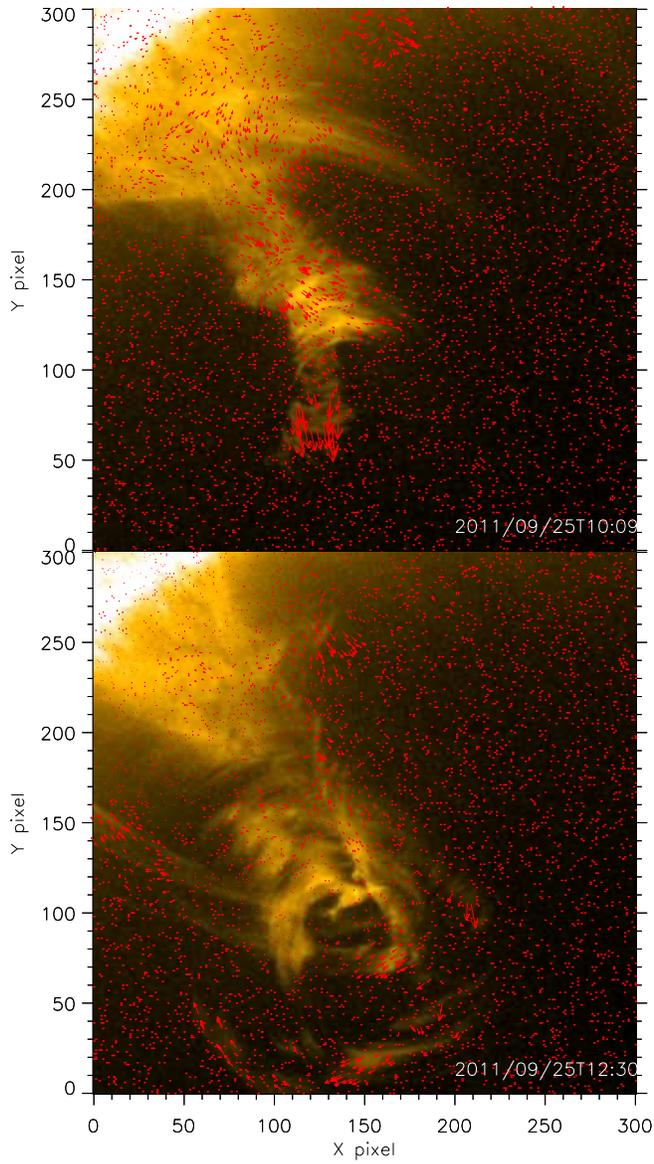}
\caption[]{Apparent velocity field computed using Local Correlation
Tracking together with AIA 171\AA\ subfield images at 2011/09/25
10:09 (top) and 2011/09/25 12:30 (bottom) of the same prominence
tornado. The maximum velocity is 55km/s at 10:09 and 95km/s at
12:30. The positive $Y$ axis points to the solar north.}
\label{velocity}
\end{figure*}

A Local Correlation Tracking (LCT) method \citep{welsch2004} is
adopted to compute the velocity field in the plane of the sky by
using two images separated by 24 seconds. Two panels in Fig.
\ref{velocity} show the velocity map at 10:09 and 12:30 with
corresponding maximum speed 55km/s and 95km/s, respectively.  Clear
parallel arcs are seen in the higher part of the structure at 10:09.
These striated patterns suggest a helical flux tube, and the
direction of movement shows that the helix is expanding upwards. The
flow gains speed substantially, even when it ascends against
gravity, suggesting that magnetic tension forces play an important
role. Another possible interpretation of the striations is of
density waves moving along a pre-existing helical structure. For the
coherent striated patterns to be apparent, the density wave must
have a wavelength close to that of the circumference of the helix
windings. A detailed model study is needed to gain further
understanding of this phenomenon. On top of the upward expansion of
the helical structure, the rotational motion of some blobs at the
top of the structure at 12:30 looks circular, and is possibly due to
material flowing along helical flux tubes. As we observe along the
axis of the helix, the apparent motion is rotational. Working from
approximate estimates of the radius of the circular motion (about
35000km for a blob of material at X=60, Y=40), and the time for a
brightness enhancement to make a complete revolution (about 3400s),
the true velocity is close to 65\kms. This is smaller than the sound
speed.


Fig. \ref{channels} shows the appearance of the tornado in four AIA channels. It
is not possible to assign a temperature for this structure directly
since the material flowing within the structure contains ions at a
large range of formation temperatures. Throughout the whole period
of tornado formation and rotation, the emission in the 304\AA\
channel (which is dominated by emission from He$^{1+}$) is almost
identical to that of the hotter 171\AA\ channel. Although the
304\AA\ channel can include emission from a hot line, the strength
of the signal suggests that the material injected into the filament
and cavity contains both hot and cold material.
The existence of cold material is also supported by the presence of
\Ha\ in the tornado structure, as shown in Fig. \ref{halpha}. The behavior in
\Ha\ is somewhat different, although the time coverage offered by
ground-based telescopes is restricted. Fig. \ref{halpha} shows that the tornado
is emitting in \Ha\, but by 15:30 (last panel) the emission comes
from the very base of the structure. This is different to the
behavior in 304\AA\ emission measured by AIA, where the whole
structure, including cavity, is bright past 15:30. It is possible
therefore that either the H becomes ionized after a few hours in the
top of the structure. 
Another factor is the sensitivity of the
ground-based observations. Certainly the Big Bear Solar Observatory
(BBSO) observations (rightmost panel) are affected by cloud later in
the day.

\section{Discussion and conclusion}
\label{discussion} At the start of the event narrow helical
structures are upwelling into the cavity. The exact source of these
helices is unclear, although it is likely that some or all of the
underlying prominence is expanding upwards into the cavity possibly
due to some disturbance at the prominence base. Alternatively, when
a helical flux tube is tightly wound, it may be unable to maintain
stability \citep{sakurai1976, hoodpriest1979, baty2001} and may
eventually expand, upwell or untwist into the surrounding cavity
without outside influence \citep{liggett1984}. Following this
initial development, the complex appearance and evolution of the
tornado can be interpreted as a combination of several different
actions:
1. The core of the tornado is formed of highly twisted magnetic
fields which are unstable and interact with the surrounding cavity,
possibly through reconnection.
  This interaction can result in sporadic localized brightenings and flows along the fields.
2.  Material flows upwards from the prominence base and some of this
material  ends up flowing along the helical fields of the prominence
and/or cavity. This movement appears as a rotation when viewed along
the helix axis.
3.  Density waves may propagate along the helical fields. 4. There
are larger-scale structural evolutions (for example, slow or rapid
expansion of helices, general large-scale movements)

\begin{figure*}[t]
\begin{center}
\includegraphics[width=14.0cm]{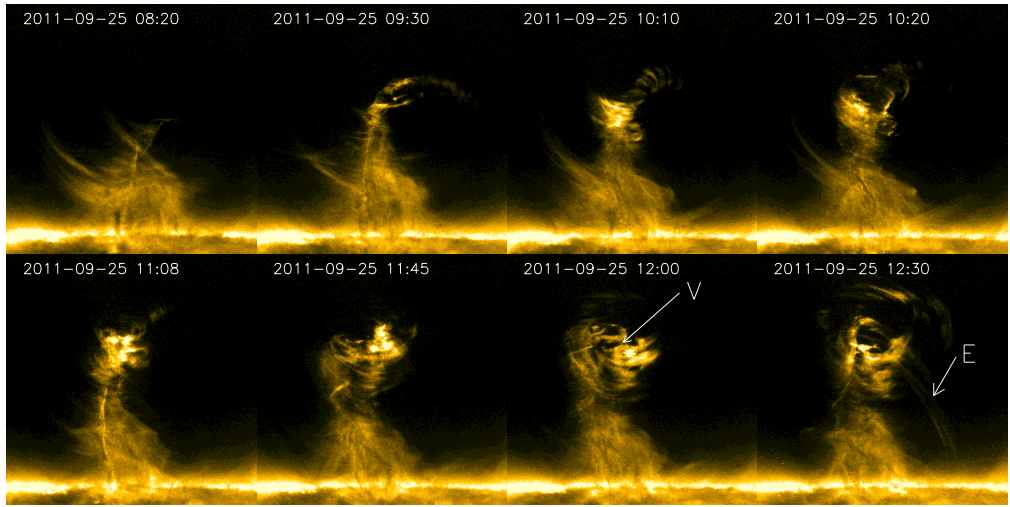}
\end{center}
\caption[]{Evolution and rotation of the tornado as seen in the AIA
171\AA\ channel over $\sim$4 hours  starting 2011/09/25 08:20 at
eight different times (see the time stamps above each frame). An
animation of this figure is available in the online version of this
journal. An event in the prominence core (possibly reconnection) is labelled `V' in the frame for
12:00, followed by an ejection labelled `E' in the following frame
for 12:30 (see text).} \label{rotation}
\end{figure*}

The filament probably consists of a highly tangled field
\citep{vanballegooijen2010}, and flow within this filament will
appear very complicated when the total emission is integrated over
the LOS. Such embedded and kinked helices will necessarily produce
favorable conditions for magnetic reconnections to occur
\citep{baty2000, kwon2008}, although high twist and reconnection do
not always lead to ejective behavior \citep{kliem2010}.
Reconnections may be difficult to observe in such a complex
structure. The apparent downflows at 12:30 seem
to originate from a region where two loop systems are seen to come
into contact at $\sim$ 12:00 (see Fig. \ref{rotation}).



\begin{figure*}[h]
\begin{center}
\includegraphics[width=13.0cm]{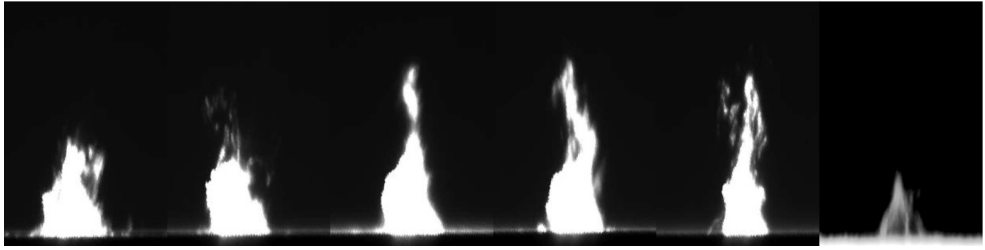}
\end{center}
\caption[]{The tornado as observed by ground-based \Ha\ telescopes.
Panels show times 07:21, 08:57, 11:09, 11:32, 13:33, 15:30 (left to
right). The observations are made by the Pic du Midi \Ha\
coronagraph except for the rightmost panel,  which is made by the
Big Bear Solar Observatory (see acknowledgments).} \label{halpha}
\end{figure*}

\citet{wang2010} studied rotation in coronal cavities and invoked a
flow of material along an arcade of loops prior to the loops
becoming detached from the solar surface and forming a helix. Their
description does not seem consistent with this event, where more
sporadic injections of material, and more rapid magnetic structure
evolution, are observed. Although occasionally the top part of the
prominence looks detached from the lower part (Fig. \ref{halpha} at
11:09), there is a continuous flow between the two parts and the
upper part is not physically detached from the lower part of the
prominence. Whether there is a preference for equatorward rotation
in such tornado-like events, as suggested by \citet{wang2010}, is a
matter for further observational study. If there is such a trend,
there must be a preferential direction to filament/cavity helicity
and a preferential direction for material flow along the structure.

Although the flow originates from a channel 14-18Mm above the solar
limb with its width $\leq 4$Mm, circular motions as wide as 90Mm are
observed at about 1.2\Rs, not much smaller than the diameter of the
cavity (roughly 110-130Mm from images of 193\AA).
Observations of the large circular motions and flows originating
from a narrow channel may shed some light to the question why a
cavity exists above a prominence. If most of the magnetic field flux
in the cavity is rooted in a small region in the lower atmosphere,
the small region may be simply unable to supply sufficient material
into the cavity unless a dramatic injection of material is caused by
some catastrophic event at the prominence base. The general
(quiescent) case would therefore be of a dark cavity devoid of
plasma due to the restrictive geometry of the flux tube at low
heights.

The dynamics and shape of this prominence and cavity are
significantly more complex than those of the erupting prominence
reported by Kurokawa et al. (1987), rotational spicules reported by
Pike \& Mason (1998), helical `EUV sprays' reported by Harrison et
al. (2001), and the emerging helical prominence reported by Okamoto
et al. (2010). The prominence is several times higher than the
emerging cool column reported by Okamoto et al. (2010), the flow is
also a few times faster and the rotation is more coherent. The fact
that the prominence reported in this paper contains plasmas at both
cool (10$^4$ K) and hot coronal temperatures while the emerging
prominence reported by Okamoto et a. (2010) is quite cool (at $\sim
10^4$K) suggests that the mechanisms which drive the flows in the
two events may be different. Recently, Berger et al. (2011) discovered coronal-temperature plasma
bubbles being injected into coronal cavities from below. Although
the bubbles are small compared to cavities, they argue that the
discovery offers an explanation for the 8-10 km s$^{-1}$ flows
observed by Doppler velocity measurements in cavities
\citet{schmit2009}.
Such quiescent prominence convection is a more gradual and
consistent process than the dynamical event described here, although
it is possible that the gradual build-up of plasma, magnetic flux
and helicity to the cavity contributed to destabilization.

This huge tornado-like structure is complex and is a compelling case
for further study. Similar dynamic events associated with a
prominence and cavity are usually expected to erupt as a coronal
mass ejection. This structure does not erupt, and remains
dynamically coherent for several hours. It is therefore an
interesting event which may shed light on the relationship between
prominences and cavities, evolution of helical fields in the low
corona, movement of material within cavities, limits on magnetic
structural stability prior to eruption, and the general structural
characteristics of cavities. Observations of such dynamical events
by AIA/SDO are placing new challenges to interpretation and models,
and will lead to a deeper understanding of the solar atmosphere.



\begin{acknowledgements} We are grateful to an anonymous referee for
the very constructive comments. The data used are provided courtesy
of NASA/SDO and the AIA science team. Images provided by  the Pic du
Midi \Ha\ coronagraph via the L`Observatoire de Paris's Bass2000
online database, and Big Bear Solar Observatory/New Jersey Insitute
of Technology have been used. This work is conducted under an STFC
grant to the Solar System Physics group at Aberystwyth University.
\end{acknowledgements}


\end{document}